# A New View to Mission Profiles

Horst Lewitschnig, PhD, Infineon Technologies Austria AG, Austria

Marcus Mayrhofer, PhD, TU Vienna, Austria

Peter Filzmoser, Prof., PhD, TU Vienna, Austria

*SUMMARY & CONCLUSIONS*

Mission profiles cover the conditions that a component, e.g., an electronic component of a vehicle, is exposed to during its lifecycle. Currently, these profiles typically provide descriptive summaries, such as histograms, of single stress parameters like temperature, humidity, or voltage.

This is highly aggregated information. New requirements for electric and autonomous driving cars require much more information how applications are used. In this work, we present a new approach for mission profiles which contains detailed usage information. We suggest a functional description over time, which allows joint modeling of various characteristics such as temperature, humidity, and voltage. The entire lifecycle history is covered, and the method can control the temporal resolution, i.e., the level of details of a mission profile. As a result, more accurate mission profiles can be generated, user quantiles can be derived, and usage outliers can be identified. This model establishes a framework to exchange usage data between suppliers, original equipment manufacturers (OEMs), and end customers while data integrity and protection are assured.

## 1 INTRODUCTION

Mission profiles are representations of the operation conditions of devices and reflect how end customers use them. Nowadays, mission profiles are typically given as histograms over temperature or temperature cycles [1]. This is high-level information about the device usage.

Requirements for electronic devices are increasing, like longer lifetimes, e.g., for components in electric and autonomous driving vehicles, or lower failure rates. To be able to design devices that fulfill these requirements, more precise knowledge about their field usage is needed. With increasing digitalization, car manufacturers get more and more application data from field usage. These application data can be used to derive advanced mission profiles. Such advanced mission profiles are highly accurate. They also reflect multiple stress factors that the devices in applications are exposed to in the field, e.g., temperature, temperature cycles, humidity, voltages, and operating hours. Accounting for the time dependency and how stress factors relate to each other is particularly important for its reliability; see, e.g., [2] for the interaction between temperature cycling and humidity stress. We base our model on multivariate functional data analysis (FDA) to create advanced mission profile models that conform to these new requirements.

Functional data analysis is a statistical approach to analyze data observed as functions over a defined domain rather than discrete multivariate vectors. It provides a method to analyze and model data with inherent temporal or spatial structures and to gain insights into multivariate functional dependencies. FDA can work with various data types like multivariate functional data, longitudinal functional data, functional time series, and spatially functional data [3, 4].

FDA is applied in various fields such as medicine, biology, economics, metrology, and engineering. Key concepts in FDA include smoothing techniques, functional principal component analysis, functional regression, functional time series analysis, and functional outlier detection. Standard resources that cover those topics are the books of [3] and [4], as well as the review papers of [7] and [8]. In this work, we apply the methods of FDA to a new application, i.e., to mission profiles.

## 2 CLASSICAL MISSION PROFILE

To produce sustainable devices that meet customer demands, the devices must withstand a large majority of the stresses they are exposed to during their lifetime. For example, a target could be that devices withstand $\alpha$ percent (e.g., $\alpha \geq 95\%$) of the stresses that occur during customers' operations. To achieve this goal, one can analyze the real-world usage demands of the device, and the product can be designed accordingly. The target level $\alpha$ is a crucial value given by the manufacturer as part of this quality policy.

Nowadays, automotive mission profiles are frequently given as histograms of temperatures [1] as illustrated in

Figure 1. Further mission profiles for temperature cycles and vibration are given in [1] as well. Several classes are defined depending on where a component is positioned in the car.

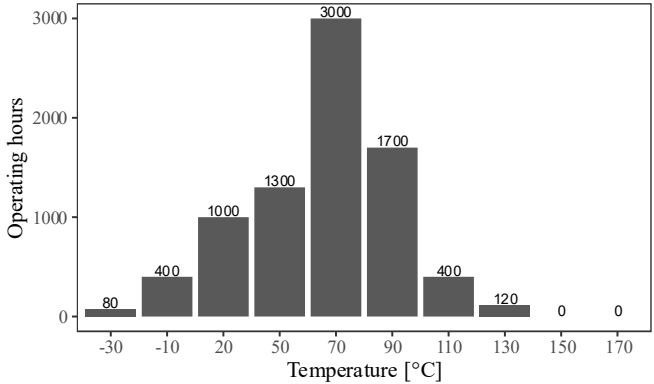

*Figure 1: Mission Profile Showing the Operating Hours at Predefined Temperatures According to [1].*

Another example is [11], where yearly mileage and survivability of cohorts are investigated. This gives a refined picture of mileage over time, and how long automotive vehicles are used.

While mission profiles in the form of histograms provide a starting point for product design, more detailed insight into the real-world usage conditions of devices is needed. To illustrate this, we consider an example where we analyze the in-field usage temperature. Let $x_i(t), t \in [0,T]$, denote the temperature of the $i$th device at time $t$, and $\boldsymbol{x}(t) = (x_1(t), \ldots, x_n(t))^T$ the temperatures of all $n$ devices at time $t$. To obtain a mission profile as shown in Figure 1, one could create a histogram based on the $\gamma$th empirical quantile of the cumulated operating temperatures up to time point $t \in [0,T]$. However, using the pointwise empirical quantile does not account for the actual data structure. The observations $x_i(t), t \in [0,T]$ are functions of time, and by using pointwise empirical quantiles the time-dependency is lost, which can lead to unreliable quantiles [10]. Hence, if the usage data are not provided in addition to the mission profile, there is no possibility to detect and address this issue.

In contrast to histogram-based mission profiles, we suggest creating mission profiles using FDA. This allows us to account for the functional nature of real-world usage data and create more accurate mission profiles. It enables us to identify, e.g., 95% of the users with the most similar usage profiles as well as usage outliers. This enables manufacturers to improve product designs by ensuring that the devices can withstand the stress they encounter throughout their lifetime, see [11], Appendix 7. Additionally, the basis of the Hilbert space can be used to aggregate and anonymize real-world usage data such that they can be shared between OEMs and their suppliers.

## 3 COMPLETE MISSION PROFILE

For each device, the complete information about the usage is given in a multivariate path which allows us to consider time-dependent information, the order of the stresses, and dependencies between stress factors. Such dependencies are, e.g., temperature and relative humidity, which depend on each other. Typically, stress observations are recorded over time. We employ a smoothing procedure with basis functions of a Hilbert space. By this, we can steer the required resolution level while establishing a mathematical framework for further processing. Using techniques from FDA, we can identify the load conditions shared by a substantial proportion of users and account for the multivariate and time-dependent relationships between the stress factors. In comparison, the commonly used histograms of the stress parameters do not capture those dependencies nor the order in which stresses occur. Advanced mission profiles based on FDA enable suppliers to tailor their device designs to the profile that represents real-world usage conditions.

With this model, we can analyze field-usage data and relate them to standardized qualification stress tests according to the respective acceleration laws.

A set of such multivariate paths can reflect the field-usage conditions of a population of devices and contains all necessary information to construct mission profiles. With FDA, we can describe and analyze these functional observations. Furthermore, we can identify quantiles and multivariate outlyingness scores. We can determine usage paths that cover, e.g., $\geq 95\%$ of the users, as well as identify anomalous usage behaviors.

## 4 MATHEMATICAL DESCRIPTION

In real operation, the lifetime of a device is influenced by various stress factors that can be monitored and assessed over time. Hence, every observed stress factor is a multivariate function of time, and FDA provides a suitable framework for this setting, allowing us to incorporate selected usage parameters into mission profiles. With this setup, we can account for the covariance structure and the stress history, incorporating valuable information that is commonly lost when using univariate summaries. Objects of interest in FDA are functions, and the underlying statistical model of those functions are stochastic processes. However, in reality we do not observe continuous functions, but data are digitalized and collected discretely over time. Therefore, a smoothing procedure is applied to the observed data as a preliminary treatment. We assume that we observe the parameters $\boldsymbol{y}_{il} = (y_{i1l}, \ldots, y_{ipl})^T \in \mathbb{R}^p$ at time points $t_l \in [0,T]$, $l = 1, \ldots, q$, where $p$ is the number of observed parameters. Further, we assume $\boldsymbol{y}_{il} = \boldsymbol{x}_i(t_l) + \boldsymbol{\varepsilon}_i$, where $\boldsymbol{x}_i \in \mathcal{H}$ is a sample of the latent stochastic process $\boldsymbol{x} \in \mathcal{H}$ we are interested in, $\mathcal{H}$ is a Hilbert space, and $\boldsymbol{\varepsilon}_i$ is a random error with an expected value of zero. The smoothing procedure aims to eliminate the contribution of the error term and obtain an approximation of $\boldsymbol{x}_i$. Using basis representation, we model $\boldsymbol{x}_i(t)$ by $\tilde{\boldsymbol{x}}_i(t)$ with coordinates

$$\tilde{x}_{ij}(t) = \sum_{k=1}^{K} c_{ijk}\phi_k(t), \quad (1)$$

where $\boldsymbol{\phi} = (\phi_1, \ldots, \phi_K)^T$ is a family of $K$ basis functions. That can be, e.g., basis splines, wavelets, or periodic functions like sine and cosine. The selection of the base functions depends on the structure of $\boldsymbol{y}_{il}$. Basis splines are a general choice, while, e.g., for high frequency data periodic functions are used.

The coefficients $c_{ijk} \in \mathbb{R}$ can be determined using penalized smoothing such that

$$\sum_{j=1}^{p}\left(\sum_{l=1}^{q}\left(y_{ijl} - \tilde{x}_{ij}(t_l)\right)^2 + \lambda \int_0^T \left(\frac{d^2}{dt^2}\tilde{x}_{ij}(t)\right)^2 dt\right) \quad (2)$$

is minimized, and instead of the second order derivative, linear combinations of derivatives of $\tilde{x}_{ij}(t)$ could be used in the penalty term. In Eqn. (2), higher $\lambda$ results in more smoothing. By choosing a proper value of $\lambda$ and the number of basis functions, we can control the smoothness of $\tilde{\boldsymbol{x}}(t)$. This allows us to control the level of granularity of the mission profile. In practice, functional data are digitalized and recorded at discrete time points and contain noise. By smoothing the data, we can capture the signal contained in the raw data and filter out the noise. Moreover, this approach still works if the observations or their individual coordinates are observed on different time grids, and it can be extended to settings where different basis functions are used for each coordinate.

The preprocessing steps allow us to transform the raw data into functions. These are elements of a Hilbert space, such as the space of $p$-dimensional square-integrable functions $\mathcal{H} \coloneqq L_2([0,T])^p = L_2([0,T]) \times \cdots \times L_2([0,T])$ endowed with inner product $\langle \cdot, \cdot \rangle_\mathcal{H}$ and norm $\|\cdot\|_\mathcal{H}$. Let $\boldsymbol{f} = (f_1, \ldots, f_p)^T$ and $\boldsymbol{g} = (g_1, \ldots, g_p)^T$ be two random functions in $\mathcal{H}$, then

$$\langle \boldsymbol{f}, \boldsymbol{g} \rangle_\mathcal{H} = \sum_{j=1}^{p} \int_0^T f_j(t) g_j(t) dt \quad (3)$$

and $\|\boldsymbol{f}\|_\mathcal{H} = \langle \boldsymbol{f}, \boldsymbol{g} \rangle_\mathcal{H}^{\frac{1}{2}}$.

The $p$-dimensional mean function of $x \in \mathcal{H}$ is given by

$$\boldsymbol{\mu}(t) = \mathbb{E}[\boldsymbol{x}(t)] = \left(\mathbb{E}[x_1(t)], \ldots, \mathbb{E}[x_p(t)]\right)^T, \quad (4)$$

and the $p$-variate covariance function is defined as

$$\boldsymbol{\Sigma}(s,t) = \begin{pmatrix} \Sigma_{11}(s,t) & \cdots & \Sigma_{1p}(s,t) \\ \vdots & \ddots & \vdots \\ \Sigma_{1p}(s,t) & \cdots & \Sigma_{pp}(s,t) \end{pmatrix} \quad (5)$$

with (cross) covariance functions

$$\Sigma_{jk}(s,t) = \mathbb{E}\left[\left(x_j(s) - \mu_j(s)\right)\left(x_k(t) - \mu_k(t)\right)\right] \quad (6)$$

for $j, k \in \{1, \ldots, p\}$. The element $\Sigma_{jk}$ of $\boldsymbol{\Sigma}$ quantifies the dependence between functions $x_j$ and $x_k$ [5]. For example, if $\Sigma_{jk}(s,t)$ is large, then $x_j(s)$ and $x_k(t)$ tend to be simultaneously below or above their respective means at those points [6]. For a more detailed overview of multivariate functional data and an outline of how to efficiently estimate the covariance function for multivariate functional data, we refer to [12]. If the covariance function $\boldsymbol{\Sigma}(s,t)$ is continuous, then the associated covariance operator $\boldsymbol{\Gamma}: \mathcal{H} \rightarrow \mathcal{H}$ with kernel $\boldsymbol{\Sigma}(s,t)$ is a compact linear operator. With $\boldsymbol{\Sigma}_j = (\Sigma_{j1}, \ldots, \Sigma_{jp})$ and $\boldsymbol{f} \in \mathcal{H}$, the $j$th element of $\boldsymbol{\Gamma f}$ can be written as

$$(\boldsymbol{\Gamma f})_j(s) = \langle \boldsymbol{\Sigma}_j(s, \cdot), \boldsymbol{f} \rangle_\mathcal{H} = \sum_{k=1}^{p} \int_0^T \Sigma_{jk}(s,t) f_k(t) dt. \quad (7)$$

In the setting of multivariate functional data, the covariance operator $\boldsymbol{\Gamma}$ takes the same role as the covariance matrix in multivariate statistics. In the multivariate setting, where the observations are discrete vectors, the Mahalanobis distance [13] could be used to determine a proportion of, e.g., $\gamma \geq 95\%$ of observations that share common characteristics. It provides a center outward ordering of the data and accounts for dependencies between parameters based on the inverse of the covariance matrix. An ordering is needed, but we refrain from this operator in the functional setting because the covariance operator $\boldsymbol{\Gamma}$ is non-invertible because it is a compact linear operator. Instead, statistical depth functions are used for outlier detection and to obtain a center outward ordering of the functions. Those were initially proposed for multivariate data to provide multivariate counterparts to univariate order statistics [14]. Relative to a distribution function $\boldsymbol{F}$ on $\mathbb{R}^p$, the depth function $D(\boldsymbol{x}, \boldsymbol{F})$ provides a center outward ordering of the samples $\boldsymbol{x} \in \mathbb{R}^p$. Observations with large depth values are more central, and the instance with the largest depth defines the center or median. Depth functions can also be related to outlyingness measures such as the Mahalanobis distance. If $O(\boldsymbol{x}, \boldsymbol{F})$ is an unbounded measure of the outlyingness of $\boldsymbol{x} \in \mathbb{R}^p$ with respect to the center of $F$, then the connected depth function can be obtained by $D(\boldsymbol{x}, \boldsymbol{F}) \coloneqq \left(1 + O(\boldsymbol{x}, \boldsymbol{F})\right)^{-1}$.

Many different definitions for depth functions exist in the functional setting. For univariate curves, band depth [15] is a prominent example, and it was used to introduce functional boxplots [10]. For multivariate functions, the multivariate skew-adjusted projection depth of [16] and the functional directional outlyingness of [17] are frequently used. A taxonomy for the different kinds of functional outliers was introduced in [16]. We distinguish between isolated outliers, which are outlying only for very short time periods, and persistent outliers, outlying for more extended time periods. We can further differentiate between magnitude, shape, and amplitude outliers in the latter category. Functional directional outlyingness measures the outlyingness of functional data by considering the level and the direction of their deviation from the central region. The magnitude-shape plot visualizes the directional and shape outlyingness and provides a tool for functional outlier detection. Such a decomposition is especially useful for multivariate functional data because every marginal function would have to be visually inspected to determine the outlier type. For a set of (smoothed) observations $\boldsymbol{x}_i, i \in \{1, \ldots, n\}$, an outlyingness measure $O(\boldsymbol{x}_i, \boldsymbol{F})$ can be used to identify the set $H \subseteq \{1, ..., n\}, |H| = \lceil n\gamma \rceil$, of observations with the lowest outlyingness. Hence, if the observations $x_i, i \in \{1, ..., n\}$, contain the usage information of $n$ devices, then the set $H$ represents the $\gamma$ percent of the devices with the most similar usage profiles defining complete mission profiles.

These complete mission profiles based on multivariate paths can also be converted back to well-known summary statistics and visualizations. Let $x_{.j}$ be a parameter like temperature or voltage. We can compute the average duration of a subset of devices for which the parameter is within a certain range of values during the whole usage duration. Let $x_{ij}(t)$ denote the temperature of the $i$th device at time $t \in [0, T]$ and $[a, b)$ the temperature range. Furthermore, we can partition the temperature range $[a, b)$ into $M$ disjoint temperature intervals $[a, b) = \bigcup_{m=1}^{M} r_m, r_m = [a_{m-1}, a_m)$, where $a_0 = a$ and $a_M = b$. Then, the average duration of the temperature between $r_m = [a_{m-1}, a_m)$ for the devices in H is given by

$$R_j(H, r_m) = \sum_{i \in H} \int_0^T I(x_{ij}(t) \in r_m) dt, \quad (8)$$

with $I(x_{ij}(t) \in r_m)$ denoting the indicator function for the temperature $x_{ij}(t)$ of device $i$ to be in the temperature region $r_m$. For other parameters, such as milage, we are interested in the parameter value at the last timepoint. Let $x_{ik}(t), t \in [0, T]$ denote the milage of the $i$th device. Then we can compute a histogram of the milage at the last timepoint $T$ of all observations that are in $H$, i.e., a histogram of $x_{ik}(T), i \in H$.

## 5 APPLICATIONS

Identifying and understanding common usage behaviors shared by a large proportion of users is a vital step in product design. In Figure 2, we illustrate this with a simplified example based on simulated daily vehicle-milage data of four user types for one year. The three groups, colored in gray, orange, and violet, represent privately owned vehicles, and we distinguish between users who drive average, long, or short distances, respectively. The last group is colored in green and represents the daily mileage of vehicles from a carsharing company. We assume that private owners randomly drive longer distances for trips and holidays, resulting in a steep increase in mileage, while the shared vehicles accumulate more mileage continuously.

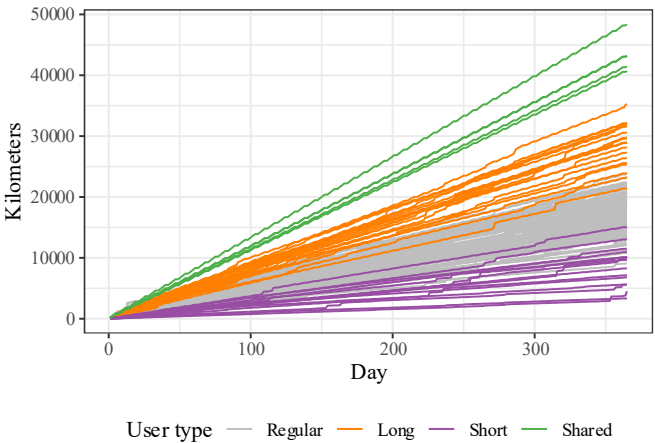

*Figure 2: Yearly Vehicle-Mileage Data of 4 Different User Groups.*

In Figure 3, we created a functional boxplot of the yearly mileage data. The black line indicates the median, and the dark gray area delimited by the dark blue lines depicts the 50% central region, i.e., the band around the half of the observations with the highest depth corresponding to the standard boxplot inter-quartile range (IQR). The outer blue lines are the fences, the band around the observations contained in the region obtained by expanding the 50% central region by 1.5 times the IQR. The light gray area shows the 95% central region. In the functional boxplot, observations that are outside of the fences at least once are flagged as outlying [10]. The outlying usage profiles in our example are colored red, and most of them correspond to the usage profiles of the shared vehicles and a few of the users who travel longer distances.

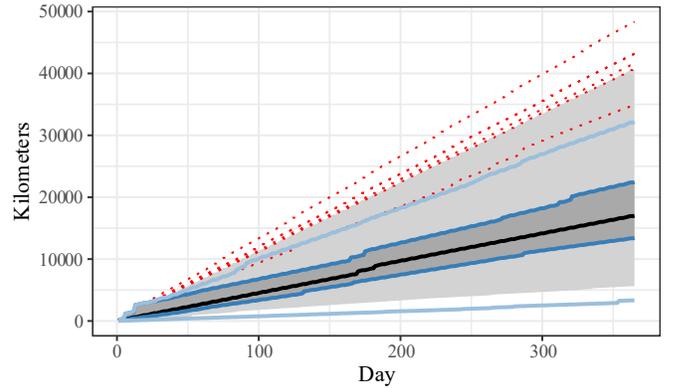

*Figure 3: Functional Boxplot of Yearly Vehicle Milage Data.*

As a second multivariate example, we consider the temperature cycles of a semiconductor installed in a car along with its mileage. While we considered aggregated daily measurements in the first example, we used a minute-wise resolution for a single day in the second example. The temperature cycles are created as follows: If the car is not used for an extended period, the semiconductor is at ambient temperature. When the vehicle is turned on, the temperature of the semiconductor gradually increases until it reaches a maximum level, maintaining this upper limit while the vehicle is in use. Finally, the semiconductor cools to ambient temperature when the car is turned off.

In Figure 4, we show the simulated data including three different usage profiles. Most observations correspond to privately operated vehicles used at an ambient temperature of 20 °C, colored in gray. Secondly, we added two usage profiles of privately operated cars located in a cold area with an ambient temperature of −30 °C, colored in blue. Lastly, we consider two usage profiles of vehicles of a carsharing company operated in an area with an ambient temperature of 20 °C which are used more frequently for shorter trips, colored in green.

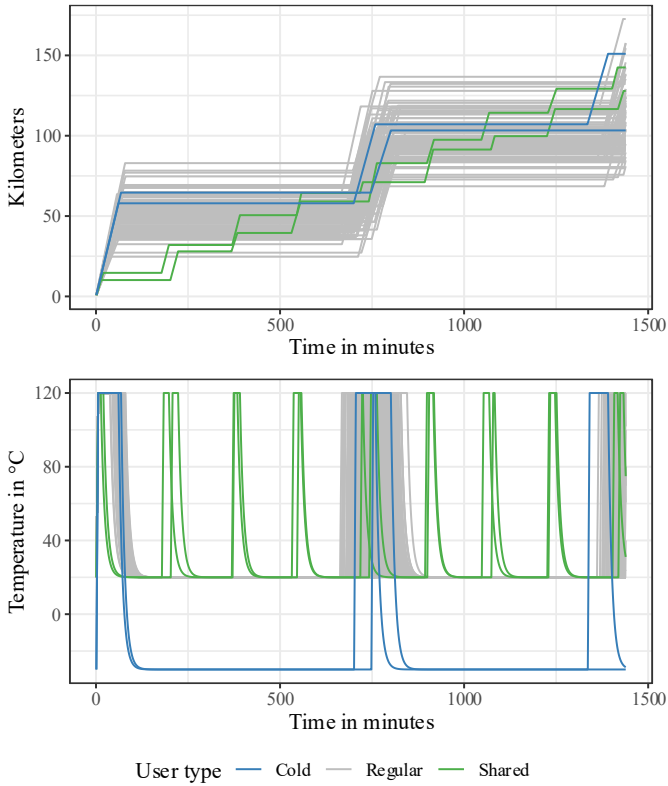

Figure 5: Simulated Daily Car Profiles of Mileage and Temperature vs. Time

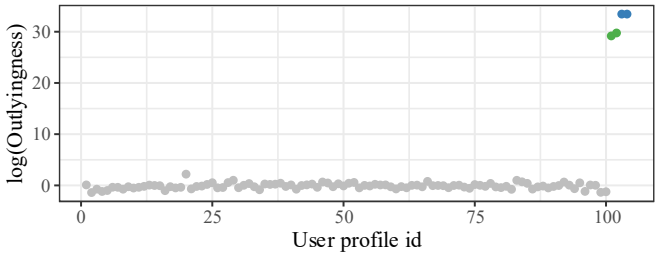

Figure 5: Functional Adjusted Outlyingness Scores of the Profiles of Figure 5. Green and Blue Points are Outliers.

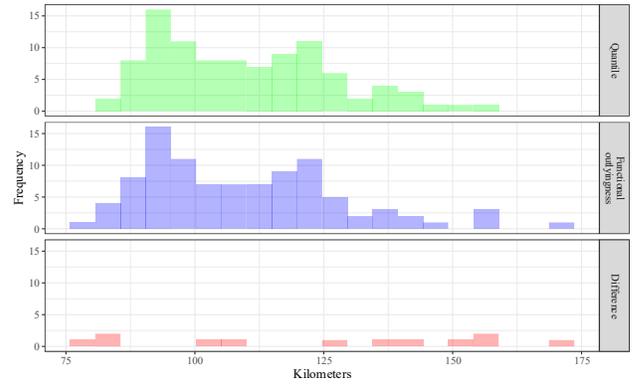

Figure 6: Histograms of Milages, Functional Outlyingness, and Their Difference.

Based on the *functional adjusted outlyingness* proposed by [16], we can identify the outlying usage profiles as illustrated in Figure 5. By default, directional outlyingness uses random projection depth, where the number of random projections controls computational complexity. We can distinguish between unusual usage profiles, corresponding to the cars used at lower ambient temperatures and the frequently used carsharing vehicles, and the regular observations based on their outlyingness, which is plotted on a log scale in this example. This method identifies outliers. It is up to the manufacturer how to deal with such outliers, e.g., to cover it in design or to exclude them.

In Figure 6, we show histograms of milage at the last time point based on the 95% most central observations according to two different criteria. The top plot shows the subset of observations between the 2.5% and 97.5% empirical quantile of the mileage at the last time point. The middle plot shows 95% of the observations with the lowest functional outlyingness along their paths over the whole timeframe, and the bottom plot shows the difference between the histograms. The identification of outlying usage profiles can be useful, e.g., for field failures, learnings from mission profiles, or warranty claims. Especially the increasing digitalization opens an opportunity to detect outlying behavior as well during operations. It could then be possible to, e.g., adapt the operation modes of such devices.

## 6 CONCLUSIONS & OUTLOOK

We propose a multivariate functional description of usage conditions, covering the full lifecycle history of components in a desired granularity. This advantage is a more refined description of mission profiles, incorporating stress history and correlations between stress factors. As a result, we can define quantiles of mission profiles that cover a predefined proportion (e.g., ≥ 95%) of all users. Furthermore, we can identify outlying usage behavior. With this model, product design can cover much more precisely the applications and market needs. Furthermore, this method can be used to determine reliability tests like for product qualification. It directly supports robustness validation according to [11]. As an outlook, mission profiles can also be used for product verification. Design points can be combined with the likelihood that they are used in real life. This can be extended to a risk criterion, which then serves as an input to the verification plan. Also, the structure of available usage data needs more investigations. If a manufacturer collects usage data, then he can only collect data from users that agreed to exchange data, which is a subset of the whole population of users. Furthermore, empirical mission profiles can contain censored data. This setting was investigated for univariate functional data in [18] and [19], and the model should be further developed to picture such incomplete data as well.

*ACKNOWLEDGEMENT*

This work is part of the AI4CSM project and has received funding from the ECSEL Joint Undertaking (JU) under grant agreement No 101007326. The JU receives support from the European Union's Horizon 2020 research and innovation programme and national authorities. This work is also funded by the Austrian IKT der Zukunft programme via the Austrian Research Promotion Agency (FFG) and the Austrian Federal Ministry for Climate Action, Environment, Energy, Mobility, Innovation and Technology (BMK) under project No 884070.